\documentclass[11pt]{article}
\usepackage{amsmath,amssymb,tikz}
\usepackage{graphicx}
 \allowdisplaybreaks
\long\def\symbolfootnote[#1]#2{\begingroup%
\def\thefootnote{\fnsymbol{footnote}}\footnote[#1]{#2}\endgroup}

\def\be{\begin{equation}}
\def\ee{\end{equation}}
\def\ba#1\ea{\begin{align}#1\end{align}}
\def\bg#1\eg{\begin{gather}#1\end{gather}}
\def\bm#1\em{\begin{multline}#1\end{multline}}
\def\bmd#1\emd{\begin{multlined}#1\end{multlined}}

\def\d{\delta}

\def\e{\epsilon}

\def\g{\gamma}

\def\m{\mu}
\def\n{\nu}

\def\r{\rho}
\def\s{\sigma}
\def\t{\tau}

\def\la{\label}

\def\re{\ref}
\def\er{\eqref}

\def\sse{\subsection}
\def\ssse{\subsubsection}
\def\fr{\frac}

\def\pa{\partial}

\def\ph{\phantom}
\def\eq{\equiv}

\def\qqu{\qquad}
\def\lt{\left}
\def\rt{\right}
\def\({\left(}
\def\){\right)}
\def\[{\left[}
\def\]{\right]}
\def\<{\langle}
\def\>{\rangle}

\def\Tr{{\rm Tr}}
\def\cO{{\mathcal O}}

\def\zb{{\bar z}}

\newcommand{\Tdu}[4]{{#1_{\ph{#2}#3}^{#2\ph{#3}#4}}}

\newcommand{\aei}{{\it Max Planck Institute for Gravitational Physics
(Albert Einstein Institute)\\ Am M\"uhlenberg 1, 14476 Golm,
Germany}}

\newcommand{\sta}{{\it Stanford Institute for Theoretical Physics,
Department of Physics, Stanford University,\\
Stanford, CA 94305, USA}}

\newcommand{\ias}{{\it School of Natural Sciences, Institute for Advanced Study,\\
Princeton, NJ 08540, USA}}
\begin{document}

\thispagestyle{empty}
\begin{flushright}
\hfill{SU-ITP-15/15}
\end{flushright}
\begin{center}

~\vspace{20pt}

{\Large\bf Generalized Gravitational Entropy from Total Derivative Action}

\vspace{25pt}

Xi Dong\symbolfootnote[1]{Email:~\sf xidong@ias.edu}, Rong-Xin Miao\symbolfootnote[2]{Email:~\sf rong-xin.miao@aei.mpg.de}

\vspace{10pt}${}^\ast{}$\sta

\vspace{10pt}${}^\ast{}$\ias

\vspace{10pt}${}^\dagger{}$\aei

\vspace{2cm}

\begin{abstract}
We investigate the generalized gravitational entropy from total derivative terms in the gravitational action.  Following the method of Lewkowycz and Maldacena, we find that the generalized gravitational entropy from total derivatives vanishes.  We compare our results with the work of Astaneh, Patrushev, and Solodukhin.  We find that if total derivatives produced nonzero entropy, the holographic and the field-theoretic universal terms of entanglement entropy would not match.  Furthermore, the second law of thermodynamics could be violated if the entropy of total derivatives did not vanish.
\end{abstract}

\end{center}

 \newpage
 \setcounter{footnote}{0}
 \setcounter{page}{1}

\tableofcontents

\section{Introduction}

A remarkable property of quantum gravity is that the gravitational entropy associated with a horizon is given by its area \cite{Bekenstein:1973ur, Bardeen:1973gs, Hawking:1974sw}:
\be\la{area}
S = \fr{\rm Area}{4G_N} \,.
\ee
The area law was shown by Gibbons and Hawking for the case of Killing horizons \cite{GH}.  It was later generalized by Ryu and Takayanagi to the holographic entanglement entropy \cite{RT}; the analog of the horizon in this case is a minimal surface and generally does not have a Killing vector.  This generalized gravitational entropy was shown to also satisfy the area law \er{area} by Lewkowycz and Maldacena \cite{Maldacena}.

It is important to note that the area law strictly applies to the case of Einstein gravity.  However, higher derivative corrections to Einstein gravity naturally arise in ultraviolet complete theories of quantum gravity such as string theory.  As we may expect, these higher derivative interactions give corrections to the area law \er{area}.  Broadly speaking, there are two types of contributions:
\be\la{swa}
S = S_{\rm Wald} + S_{\rm anomaly} \,.
\ee
The first is the Wald entropy \cite{Wald}
\begin{eqnarray}\label{Waldentropy}
S_{\text{Wald}}=-2\pi\int d^dy \sqrt{g}\frac{\delta L}{\delta R_{\mu\nu\rho\sigma}}\epsilon_{\mu\nu}\epsilon_{\rho\sigma}.
\end{eqnarray}
The second type of contributions involves the extrinsic curvature and have an anomaly-like origin \cite{Solodukhin, Dong, Camps}.  Both types of contributions may be derived by studying Euclidean conical geometries and their regularized versions.

It is well-known that total derivative terms in a gravitational action do not contribute to the equation of motion.  One might wonder whether they have any physical effect on the theory at all, and in particular, whether they contribute to the gravitational entropy.  In this paper, we use the Lewkowycz-Maldacena method \cite{Maldacena} to investigate the contribution to the gravitational entropy from total derivative terms in the action and find it to be zero.  This is in contrast with the result obtained in recent papers by Astaneh, Patrushev and Solodukhin (APS) \cite{Astaneh,Astaneh2}, where they find nonzero contributions to the gravitational entropy.  As we will explain in detail, the main differences between the two methods are whether the regularized cone approaches the singular cone away from the conical singularity, and whether the on-shell action of the singular cone is properly subtracted from that of the regularized cone.

Before proceeding, let us point out that for the case of a Killing horizon, the second term in \er{swa} which involves the extrinsic curvature vanishes, and we are left with the Wald entropy.  The contribution from total derivative terms to the Wald entropy is by definition zero, since the prescription for $\d L/\d R_{\m\n\r\s}$ in \er{Waldentropy} involves integration by parts.

The paper is organized as follows. In Sec.~\re{sec:boxr}, we review the Lewkowycz-Maldacena method and use it to study the simplest example of a total derivative term in the gravitational action: $\Box R$.  We calculate the gravitational entropy in this example by several different methods. In Sec.~\re{sec:general}, we investigate the gravitational entropy from general total derivative terms by using the techniques developed in \cite{Dong,Miao}.  We compare our method with the one used by APS in Sec.~\re{sec:compare}, and point out other problems for their results in Sec.~\re{sec:problems}.  In particular, their results imply that holographic entanglement entropy in conformal field theories does not agree with field-theoretic predictions. Moreover, we show that the second law of thermodynamics can be violated if the gravitational entropy from total derivatives is nonzero.  We give a short conclusion in Sec.~\re{sec:conclude}.

\section{Simplest example: $\Box R$}\la{sec:boxr}

In this section, we consider the simplest example of a total derivative term in the gravitational action: $\Box R$.  We first review the Lewkowycz-Maldacena method of calculating the generalized gravitational entropy \cite{Maldacena}, with special emphasis on the roles of singular and regularized cones.  We then apply it to the $\Box R$ example, and find that such a total derivative term does not contribute to the generalized gravitational entropy.

\subsection{Review of the Lewkowycz-Maldacena method}\la{sec:lm}
We start with the replica trick for calculating the generalized gravitational entropy:
\be\la{sdef}
S = -\lim_{n\to1} \fr{1}{n-1} \(\log Z_n -n\log Z_1\) \,,
\ee
where $Z_n$ is the partition function of the $n$-fold branched cover of the original Euclidean geometry.  Using the AdS/CFT correspondence \cite{Maldacena:1997re, Gubser:1998bc, Witten:1998qj}, we construct the dual bulk geometries $B_n$ and find
\be\la{son}
S = \lim_{n\to1} \fr{1}{n-1} \(I[B_n] -nI[B_1]\) \,.
\ee
It is important to remember that $B_n$ (which is defined only for integer $n \ge 1$) is not a singular bulk geometry -- it is required by the prescription of AdS/CFT to satisfy all bulk equations of motion and is therefore smooth.

To perform the analytic continuation to non-integer $n$ and ultimately take the $n\to1$ limit, we assume that the bulk geometry $B_n$ has a $Z_n$ replica symmetry which allows us to take the orbifold $B_n/Z_n \eq \hat B_n$.  The orbifold is a singular (i.e. not regularized) cone with conical deficit $2\pi (1-\fr{1}{n}) \eq 2\pi\e$ on a codimension-2 surface consisting of the fixed points of the $Z_n$ symmetry, and the generalized gravitational entropy can be calculated in terms of the on-shell action of this cone:
\be\la{eei}
S = \lim_{n\to1} \fr{n}{n-1} \(I[\hat B_n] -I[B_1]\) \,.
\ee
As emphasized in \cite{Maldacena}, at this stage we should not include any contribution from the conical singularity in the on-shell action $I[\hat B_n]$; in particular, we should not include any delta-function contribution or Gibbons-Hawking-York boundary term.  The correct prescription is to simply integrate the Lagrangian until we reach the conical singularity.  The justification for this prescription is that there are no such contributions in the on-shell action of the parent space $B_n$ as it is completely smooth.

The singular cone geometry $\hat B_n$ can easily be analytically continued to non-integer $n$ by continuously tuning the conical deficit $2\pi (1-\fr{1}{n})$.  The precise prescription is to solve all bulk equations of motion while imposing the conical deficit as a boundary condition.  This is equivalent, at least for Einstein gravity and several classes of higher derivative gravity \cite{Dong}, to inserting an appropriate cosmic brane and solving all equations of motion.  The cosmic brane is an auxiliary tool for finding the conical geometry $\hat B_n$ and does not contribute to the on-shell action $I[\hat B_n]$.

Until now we have only used singular cones in the formalism.  Where do regularized cones come into this story?  They come because the singular cone geometry $\hat B_n$ used in \er{eei} is not easy to compute for general $n$.  Even close to $n\approx1$, solving for $\hat B_n$ to linear order in $n-1$ is equivalent to solving the backreaction of a cosmic brane.  It is important to distinguish this ``on-shell'' method with an ``off-shell'' method\footnote{Such an off-shell method is appropriate on the field theory side, e.g. in the calculation of the universal part of entanglement entropy in even-dimensional CFTs via their Weyl anomaly.} which simply inserts a conical deficit without modifying the geometry away from the conical singularity.

Fortunately, we do not have to solve for the singular cone $\hat B_n$ to evaluate \er{eei}.  This is because the first-order variation of an on-shell action is purely a boundary term.  We may either calculate this boundary term directly, or use a regularized cone $\hat B_{n,reg}$ which is defined to be a smooth geometry that approaches the singular cone $\hat B_n$ sufficiently fast away from the conical singularity.  The precise meaning of ``sufficiently fast'' will become clear momentarily\footnote{One regularization that definitely approaches the singular cone fast enough is to use a smooth function with compact support so that the regularized cone becomes identical to the singular cone outside some finite radial distance away from the conical singularity.}.  Using the regularized cone, we may trivially rewrite \er{eei} as
\be\la{eeit}
S = \lim_{n\to1} \fr{n}{n-1} \[\(I[\hat B_{n,reg}] -I[B_1]\) - \(I[\hat B_{n,reg}] - I[\hat B_{n}]\) \] \,.
\ee
Now, the first term $I[\hat B_{n,reg}] -I[B_1]$ is the variation of an on-shell action; the first-order variation in $n-1$ therefore vanishes because $B_1$ satisfies all equations of motion and the regularized cone $\hat B_{n,reg}$ is by definition smooth everywhere\footnote{There are no boundary terms at the asymptotic boundary because $B_1$ and $\hat B_n$ satisfy the same boundary conditions, and the regularized cone $\hat B_{n,reg}$ is defined to approach the singular cone $\hat B_n$ fast enough so as to satisfy the same boundary conditions.}.  Therefore \er{eeit} simplifies to
\be\la{eeir}
S = -\lim_{n\to1} \fr{n}{n-1} \(I[\hat B_{n,reg}] - I[\hat B_{n}]\) \,.
\ee
The advantage of \er{eeir} over \er{eei} is that the contribution is now manifestly localized near the conical defect, as the regularized cone by definition approaches the singular cone away from the conical singularity.  Therefore \er{eeir} allows us to focus on metric expansions near the conical singularity.

\sse{Trivial entropy from $\Box R$}

In this section we show that a $\Box R$ term in the gravitational action does not contribute to the generalized entropy.  We use three different methods.  The first involves directly evaluating the contribution to the on-shell action of the singular cone in \er{eei}, and uses the total derivative to reduce its contribution to a potential boundary term.  The second method is similar but uses the regularized cone and \er{eeir}.  The third method also uses the regularized cone but calculates the integrals in \er{eeir} by brute force.

Before proceeding, let us write down the general metric of the singular or regularized cone in a coordinate system adapted to a neighborhood of the conical singularity \cite{Dong}:
\bm\la{met}
ds^2 = e^{2A} \[dz d\zb + e^{2A} T (\zb dz-z d\zb)^2 \] + \(g_{ij} + 2K_{aij} x^a + Q_{abij} x^a x^b\) dy^i dy^j \\
+ 2i e^{2A} U_i \(\zb dz-z d\zb\) dy^i + \cdots \,.
\em
Here $x^a \in \{z,\zb\}$ denotes orthogonal directions to the conical singularity, and $y^i$ denotes parallel directions.  The warp factor is
\be
A = -\fr{\e}{2} \log (z\zb) \,, \qqu
\e \eq 1-\fr{1}{n} \,,
\ee
for the singular cone $\hat B_n$.  The form of the metric \er{met} is constrained by the regularity and $Z_n$ replica symmetry of the parent space $B_n$ when $n$ is a positive integer.  The most general coefficient functions $T$, $Q_{abij}$, and $U_i$ allowed by regularity can be written as Taylor expansions\footnote{These expansions are ultimately determined by solving the bulk equations of motion, although we do not need to know the detailed solution for our current purpose.  Note that $K_{aij}$ seems also allowed by the regularity of $B_n$ (when $n$ is an integer) to have such an expansion in $r^{2\e}$, but for $1<n<2$ this would lead to a singular Ricci scalar and is therefore forbidden by Einstein's equations with a bounded stress tensor.  For higher derivative gravity such an expansion could be allowed for $K_{aij}$ as shown in \cite{Camps:2014voa} but does not affect our conclusion that the entropy from total derivatives vanishes.} in $e^{-2A}=r^{2\e}$ where $r\eq|z|$.  The first terms in such expansions are
\bg
T = T^{(0)} + \cO(r^{2\e}) \,,\qqu
U_i = U_i^{(0)} + \cO(r^{2\e}) \,,\\
Q_{z\zb ij} = e^{-2\e} Q_{z\zb ij}^{(0)} + \cO(r^0) \,,\qqu
Q_{zzij} = Q_{zzij}^{(1)} + \cO(r^{2\e}) \,,\qqu
Q_{\zb\zb ij} = Q_{\zb\zb ij}^{(1)} + \cO(r^{2\e}) \,.
\eg
We have kept the metric to sufficiently many orders in the radial expansion around the conical singularity for the $\Box R$ example.  The Ricci scalar near the conical singularity is
\be\la{rcone}
R = R_\Sigma + (1-\e)^2 \(24T^{(0)} -8\Tdu Q{(0)}{z\zb i}i +16 U_i^{(0)} U^{(0)i}\) + \cO(r^{2\e}) \,,
\ee
where $R_\Sigma$ is the intrinsic Ricci scalar of the conical surface.

We may also describe the regularized cone $\hat B_{n,reg}$ by a metric of the form \er{met}.  A simple choice of the regulator is to replace the warp factor by
\be
A = -\fr{\e}{2} \log (z\zb+b^2) \,,
\ee
where $b$ is a small positive number.

\ssse{Direct method}

Let us directly use \er{eei} to calculate the contribution of $\Box R$ to the gravitational entropy.  For this purpose we calculate the on-shell action of the singular cone.  The contribution to the on-shell action from a total derivative term such as $\Box R$ is a boundary term:
\be\la{ising}
I[\hat B_n] = \int dr d\t d^dy \sqrt{G} \Box R = -\int d\t d^d y \sqrt{g} \lim_{r\to0} r\pa_r R \,.
\ee
We should evaluate the $r\to0$ limit of the above expression for finite $n-1$, and only take the $n\to1$ limit in \er{eei} at the end.  For finite $n-1$ or equivalently finite $\e$, we find from \er{rcone} that $r\pa_r R = \cO(r^{2\e})$ and therefore \er{ising} vanishes.

To complete the calculation of \er{eei} we also need to evaluate $I[B_1]$.  Since $B_1$ is a completely smooth geometry that satisfies the bulk equations of motion, its Ricci scalar is finite and has a Taylor expansion near $r=0$.  Therefore $I[B_1]$, which also reduces to a boundary term from integrating $\Box R$, vanishes.  It is now clear from \er{eei} that $\Box R$ does not contribute to the gravitational entropy.

Note that we did not include boundary terms at the asymptotic infinity when integrating $\Box R$ in \er{ising}.  This is because such boundary terms, if nonzero, would have to be compensated by additional boundary terms at the asymptotic infinity in the gravitational action\footnote{For $\Box R$ the additional boundary action required by a well-posed variational principle is $I_{\rm bdy} = -\int_\pa d^{d+1}x \sqrt{\g} n^\m \pa_\m R$ where $\g$ is the determinant of the induced metric and $n^\m$ is outward-pointing unit normal vector.}, as required by a well-posed variational principle \cite{Dyer:2008hb}.

The reason why a total derivative term such as $\Box R$ does not contribute to the gravitational entropy is particularly clear if we first study Renyi entropies $S_n$ at integer $n>1$, defined by the expression in \er{sdef} before taking the $n\to1$ limit:
\be
S_n = -\fr{1}{n-1} \(\log Z_n -n\log Z_1\) \,.
\ee
In terms of the on-shell action of dual bulk geometries, the Renyi entropy may be written as
\be
S_n = \fr{1}{n-1} \(I[B_n] -nI[B_1]\) \,,
\ee
analogous to \er{son}.  For any positive integer $n$, the bulk geometry $B_n$ is completely smooth and the contribution to the on-shell action $I[B_n]$ from a total derivative term vanishes identically.  There is no boundary term from integrating a total derivative term at $r=0$ because $B_n$ is regular there, and the reason for the absence of a boundary term at the asymptotic infinity is the same as argued above.  Therefore all Renyi entropies $S_n$ vanish at integer $n>1$, and by analytic continuation this statement holds for all $n$, including the case of $n=1$ which gives the gravitational entropy.

\subsubsection{Boundary method}

Now let us investigate the entropy from $\Box R$ by using \er{eeir}. In this approach, we need to calculate the action difference between the regularized cone and the singular cone. We firstly integrate the total derivative to get a boundary term and then derive the entropy from this boundary term. 
For simplicity, we focus on the following regularized conical metric 
 \begin{eqnarray}\label{coneboxRDong}
ds^2=\frac{1}{(r^2+b^2)^{1-\frac{1}{n}}}(dr^2+r^2d\tau^2)+(\delta_{ij}+2r \sin \tau K_{1 ij}+2r \cos \tau K_{2 ij})dy^i dy^j \,,
\end{eqnarray}
with $\tau\sim \tau+2\pi$.  Here we have set $T=U_i=Q_{abij}=0$ in the language of \er{met}. The approach below can easily be applied to the general metric \er{met}.  By dimensional analysis,  we notice that only the $K^2$ terms contribute to the entropy. Focus on such terms, we have
\begin{eqnarray}\label{boundaryBoxRDong}
\int_0^{r_0}dr\int_0^{2 \pi}d\tau\sqrt{G}\Box R&=& \lt.\int_0^{2 \pi}d\tau\sqrt{G}G^{rr}\partial_r R \rt|^{r= r_0}_{r=0}\nonumber\\
&=&\lt. \frac{4 \pi  (n-1) \( \[3 \Tr K^2- (\Tr K)^2\]r^6+ d_1 b^2r^4+d_2 b^4 r^2\)}{ n \left(r^2+b^2\right)^{\frac{1}{n}+2}} \rt|^{r=r_0}_{r=0} \\
&=& 4\pi \[3 \Tr K^2- (\Tr K)^2\](n-1)+O[(n-1)^2] \,,\label{boundaryBoxRDong1}
\end{eqnarray}
where $d_n$ are coefficients irrelevant for the gravitational entropy. To derive \er{boundaryBoxRDong1} from \er{boundaryBoxRDong}, we have used the fact that $b\ll r_0$. According to \er{eeir}, we should subtract off the contribution of the singular cone ($b=0$). From \er{boundaryBoxRDong} and \er{boundaryBoxRDong1}, we find
\begin{eqnarray}\label{boundaryBoxRDong2}
\int_0^{r_0}dr\int_0^{2 \pi}d\tau\sqrt{G}\Box R- (b=0)=O[(n-1)^2].
\end{eqnarray}
Note that we take $n>1$ and $b$ finite for the regularized cone $\hat B_{n,reg}$, while we have $n>1$ and $b=0$ for the singular cone $\hat B_{n}$.
It is now clear that the entropy from $\Box R$ is zero by using this ``boundary method.''

\ssse{Bulk method}

Now let us use a different method to derive the entropy from $\Box R$. Instead of considering the boundary terms, we calculate the integrals in \er{eeir} by brute force. 

Similar to the above section, we take the regularized conical metric \er{coneboxRDong} and focus on the $K^2$ terms in the action. We have
\begin{align}\label{bulkBoxRDong}
&\int_0^{r_0}dr\int_0^{2 \pi}d\tau \sqrt{G} \Box R -(b=0)\nonumber\\
=&\int_0^{r_0}dr\frac{8 \pi  (n-1) \big{(} [3 \Tr K^2- (\Tr K)^2]b^6 r+ [14 \Tr K^2-6 (\Tr K)^2]b^4r^3+ [(\Tr K)^2- \Tr K^2]b^2r^5\big{)}}{ \left(r^2+b^2\right)^{\frac{1}{n}+3}}\\
&+\int_0^{r_0}dr\frac{8 \pi  (n-1)^2 \big{(} [3 \Tr K^2- (\Tr K)^2]r^7+ c_1 b^2r^5+c_2 b^4r^3+ c_3 b^6r\big{)}}{ \left(r^2+b^2\right)^{\frac{1}{n}+3}}+O[(n-1)^3]-(b=0)\label{bulkBoxRDong1}\\
=&O[(n-1)^2] \,.\label{bulkBoxRDong2}
\end{align}
Here $c_n$ are coefficients irrelevant for the result.  Eq.~(\ref{bulkBoxRDong}) contributes to the Wald-like entropy while eq.~(\ref{bulkBoxRDong1}) contributes to the ``anomaly'' part of the entropy. Naively eq.~(\ref{bulkBoxRDong1}) is of order $O[(n-1)^2]$. However, it becomes of order  $O(n-1)$ after the integration (with regularization). In the above derivation, we have used the following formulae
\begin{eqnarray}\label{ABoxRDongWald}
&&\int_0^{r_0}dr \frac{b^2r^5}{\left(r^2+b^2\right)^{\frac{1}{n}+3}}- (b=0)=\frac{1}{6}+O(n-1)\nonumber\\
&&\int_0^{r_0}dr \frac{b^4r^3}{\left(r^2+b^2\right)^{\frac{1}{n}+3}}- (b=0)=\frac{1}{12}+O(n-1)\nonumber\\
&&\int_0^{r_0}dr \frac{b^6r}{\left(r^2+b^2\right)^{\frac{1}{n}+3}}- (b=0)=\frac{1}{6}+O(n-1)\nonumber\\
&&\int_0^{r_0}dr \frac{r^7}{\left(r^2+b^2\right)^{\frac{1}{n}+3}} - (b=0)=-\frac{1}{2(n-1)}+O[(n-1)^0]
\end{eqnarray}
Note again that we take $n>1$ and $b$ finite for the regularized cone, while we have $n>1$ with $b=0$ for the  singular cone.
Using the above formulae, we derive eq.~(\ref{bulkBoxRDong2}) and find that the entropy from $\Box R$ is zero by using this ``bulk method.''

\section{Trivial entropy from total derivative action}\la{sec:general}

In this section, we investigate the generalized gravitational entropy from total derivative terms in the action by applying the method of \cite{Dong}. We find that the entropy from a general covariant total derivative action vanishes.  Similarly, the entropy from a topological invariant (i.e. a total derivative locally, but not globally) such as Lovelock gravity \cite{Lovelock:1971yv, Lovelock:1970} in critical dimensions is another topological invariant \cite{Jacobson:1993xs, Dong}.  We start by reviewing the derivation of generalized gravitational entropy for the most general higher derivative gravity \cite{Miao} and then calculate the entropy from several total derivative actions.

\subsection{Entropy for the most general higher derivative gravity} 

In this section, we briefly review the derivation of holographic entanglement entropy (HEE) for the most general higher derivative gravity $L(g, R, \nabla R, \nabla^2R, \cdots)$ following \cite{Dong,Miao}. 

Let us start with the regularized conical metric \cite{Dong,Camps}
\be\label{cone}
ds^2=e^{2A}[dzd\bar{z}+T(\bar{z}dz-zd\bar{z})^2]+2i V_i(\bar{z}dz-zd\bar{z})dy^i
+( g_{ij}+Q_{ij})dy^idy^j \,,
\ee
where $g_{ij}$ is the metric on the transverse space and is independent of $z, \bar{z}$. $A=-\frac{\epsilon}{2}\log(z\bar{z}+b^2)$ is regularized warp factor. $T, V_i, Q_{ij}$ are defined as \cite{Miao,Miao2}\footnote{We expand the conical metric in powers of $(r^2,r^n e^{\pm i n\tau})$ but not $r^{2(n-1)}$. As a result, there is a lower bound for $m$ in the expansions of $T, V, Q$ in (\ref{TVQ}). The powers of $r^{2(n-1)}$ are not forbidden by regularity (of the parent space at integer $n$). But they may change the entropy formula of the curvature-squared gravity, which leads to the violation of the second law of thermodynamics \cite{Wall2}. For this reason, we do not include powers of $r^{2(n-1)}$ in the  expansions (\ref{TVQ}). It should be mentioned that even if we included powers of $r^{2(n-1)}$ the entropy from covariant total derivatives would still vanish.}
\begin{eqnarray}\label{TVQ}
&&T=\sum_{n=0}^{\infty}\sum_{m=0}^{P_{a_1\cdots a_n}+1} e^{2 m A }T_{m\ a_1\cdots a_n}x^{a_1}\cdots x^{a_n}\,,\nonumber\\
&&V_i=\sum_{n=0}^{\infty} \sum_{m=0}^{P_{a_1\cdots a_n}+1} e^{2 m A }V_{m\ a_1\cdots a_n i}x^{a_1}\cdots x^{a_n}\,,\nonumber\\
&&Q_{ij}=\sum_{n=1}^{\infty} \sum_{m=0}^{P_{a_1\cdots a_n}} e^{2 m A }Q_{m\ a_1\cdots a_n ij}x^{a_1}\cdots x^{a_n}\,.
\end{eqnarray}
Here $z,\bar{z}$ are denoted by $x^{a}$ and $P_{a_1\cdots a_n}$ is the number of pairs of
$z,\bar{z}$ appearing in $x^{a_1}\cdots x^{a_n}$. For example, we have $P_{zz\bar{z}}=P_{z\bar{z}z}=P_{\bar{z}zz}=1$, $P_{z\bar{z}z\bar{z}}=2$, and $P_{zz\cdots z}=0$.
Expanding $T,V,Q$ to the first few terms in the notations of \cite{Dong}, we have
\begin{eqnarray}\label{TVQ1}
&&T=T_0+e^{2A}T_{1}+O(x) \,,\nonumber\\
&&V_i=U_{0 \ i}+e^{2A}U_{1\ i}+O(x) \,,\nonumber\\
&&Q_{ij}=2K_{aij}x^a+Q_{0\ abij}x^ax^b+2e^{2A}Q_{1\ z\bar{z}ij}\ z \bar{z}+O(x^3) \,.
\end{eqnarray}
According to \cite{Miao, Miao2}, $T_0, U_{0\ i}, Q_{0\ abij}$ must be functions of the extrinsic curvature tensor in order to be consistent with Wald entropy in stationary spacetime. Note that $U_0\sim K$ and it is impossible to express $U_{0\ i}$ in terms of $K_{aij}$. Thus a natural choice of $U_{0\ i}$ would be zero. In principle, the exact expressions of $T_0$ and $Q_{0\ abij}$ can be derived by using the equation of motion. It is unnecessary to derive exact expressions of $T_0$ and $Q_{0\ abij}$ in the present paper. As we shall show, the entropy of covariant total derivative terms is zero for arbitrary $T_0$ and $Q_{0\ abij}$.

Using the conical metric (\ref{cone}), we can calculate the regularized cone action $I_{\rm reg}$ as well as the singular cone action $I_{\rm singular}$ in the most general higher derivative gravity and then select the relevant terms to derive HEE: 
\begin{eqnarray}\label{HEE}
S=-\partial_{\epsilon} \(I_{\rm reg}-I_{\rm singular}\)|_{\epsilon=0} \,.
\end{eqnarray} 
Let us list all the relevant terms of HEE below \cite{Miao}. 

First class: generalized Wald entropy
\begin{eqnarray}\label{dA}
\int dz d\bar{z} z^m\bar{z}^n\partial_z^{m+1}\partial_{\bar{z}}^{n+1}A&=&(-1)^{m+n+1} m!n! \pi \epsilon \,.
\end{eqnarray}
 Equivalently, we have
 \begin{eqnarray}\label{dAD}
\partial_z^{m+1}\partial_{\bar{z}}^{n+1}A=-\pi \epsilon \partial_z^{m}\partial_{\bar{z}}^{n}\bar{\delta}(z,\bar{z}) \,.
\end{eqnarray}
Here  the delta function is defined as $\int dz d\bar{z} \bar{\delta}(z,\bar{z})=1$. We call the entropy relevant to this class as the generalized Wald entropy. In addition to the usual Wald entropy, corrections from $K_z, Q_{zz}, T_z,V_z \cdots$ (but not $Q_{z\bar{z}}, T_{z\bar{z}}, V_{z\bar{z}}, \cdots$) may appear in the generalized Wald entropy. For example, the generalized Wald entropy for action $L(g, R, \nabla R)$ is \cite{Miao}
\begin{eqnarray}\label{GWaldentropy}
S_{\text{G-Wald}}&=&2\pi\int d^dy \sqrt{g}\big{[}\ \frac{\delta L}{\delta R_{z\bar{z}z\bar{z}}}
+2(\frac{\partial L}{\partial \nabla_zR_{\bar{z}i\bar{z}j}}  K_{\bar{z}ij}+c.c ) \ \big{]}\nonumber\\
&=&2\pi\int d^dy \sqrt{g}\big{[}\ -\frac{\delta L}{\delta R_{\mu\nu\rho\sigma}}\epsilon_{\mu\nu}\epsilon_{\rho\sigma}
+2\frac{\partial L}{\partial \nabla_{\alpha}R_{\mu\rho\nu\sigma}}  K_{\beta\rho\sigma} (n^{\beta}_{\ \mu}n_{\alpha\nu}-\epsilon^{\beta}_{\ \mu}\epsilon_{\alpha\nu})\ \big{]} \,.
 \end{eqnarray}
It reduces to the usual Wald entropy for stationary black holes. Thus it is consistent with Wald's results. It should be mentioned that, due to these corrections, the generalized Wald entropy from total derivative terms is nonzero in the general case. 

Second class: anomaly-like entropy
\begin{eqnarray}\label{dAdA}
\int dz d\bar{z} z^m\bar{z}^n\partial_z^{m+1}A\partial_{\bar{z}}^{n+1}Ae^{-\beta A}&=&(-1)^{m+n+1}m!n!\frac{\pi \epsilon}{\beta} \,.
\end{eqnarray}
Equivalently, we have
 \begin{eqnarray}\label{dAdAD}
\partial_z^{m+1}A\partial_{\bar{z}}^{n+1}Ae^{-\beta A}=-\frac{\pi \epsilon}{\beta}\partial_z^{m}\partial_{\bar{z}}^{n}\bar{\delta}(z,\bar{z}) \,.
\end{eqnarray}
These terms contribute to the anomaly-like entropy. They are the would-be logarithmic terms which could gain a $1/\epsilon$ enhancement after the regularized integral.

Let us briefly discuss the proof of the above key formulas eqs.~(\ref{dA}-\ref{dAdAD}). Eqs.~(\ref{dA}, \ref{dAD}) are derived from of the well-known identity  $\partial_z\partial_{\bar{z}}A=-\pi \epsilon \bar{\delta}(z,\bar{z}).$ As for the proof of eqs.~(\ref{dAdA}, \ref{dAdAD}), one can follow the method of \cite{Dong}. Here we provide a schematic derivation. Recall that $A=-\frac{\epsilon}{2}\log(z\bar{z})$, we have  $z^m\partial_z^{m+1}A=-\frac{\epsilon}{2}(-1)^m\frac{m!}{z}$. Thus we can derive
\begin{eqnarray}\label{dAdA1}
\int r dr z^m\bar{z}^n\partial_z^{m+1}A\partial_{\bar{z}}^{n+1}Ae^{-\beta A}&=&\int  dr(-1)^{m+n}m!n!\frac{\epsilon^2}{4}r^{-1+\beta \epsilon}\nonumber\\
&=&(-1)^{m+n}m!n!\frac{\epsilon}{4\beta}r^{\beta \epsilon}|^{\infty}_{0}\nonumber\\
& \cong &(-1)^{m+n+1}\frac{\epsilon}{4\beta} m!n! \,.
\end{eqnarray}
Here $z=r e^{i \tau}$ and $\cong$ denotes equivalence after regularization. Since the conical singularity is located at $r=0$, we have ignored the contributions at $r=\infty$ in the above derivation. One can check that terms at $r=\infty$ can be removed by using suitable regularization. It should be stressed that the coefficient of a would-be log divergence $r^{\beta \epsilon}$ is universal and independent of the regularization. That is the reason why one can read off the last line of eq.~(\ref{dAdA1}) in a straightforward way without applying any specific regularization. 

Using eqs.~(\ref{cone}-\ref{dAdAD}), we can derive the entropy of the most general higher derivative gravity. 

\subsection{Trivial entropy for total derivatives}

Let us compute the entropy for the following list of total derivative terms \be
\{\Box R, \Box R^2, \Box (R_{\mu\nu}R^{\mu\nu}), \Box (R_{\mu\nu\rho\sigma}R^{\mu\nu\rho\sigma})\}
\ee
by applying the method of the above section. For simplicity, we set $V_i=0$ for the squashed cone (\ref{cone})\footnote{This is also the case investigated in \cite{Astaneh,Astaneh2}.}. It should be mentioned that our metric for the regularized cone (\ref{cone}) is different from the one used in \cite{Solodukhin}, in that our regularized cone approaches the singular cone away from the conical singularity.

Let us start with the regularized conical metric (\ref{cone}) with $T, V, Q$ given by
\begin{eqnarray}\label{QijDong}
&&T=T_0+e^{2A}T_{1}+O(x) \,,\nonumber\\
&&V_i=0 \,,\nonumber\\
&&Q_{ij}=2K_{aij}x^a+Q_{0\ abij}x^ax^b+2e^{2A}Q_{1\ z\bar{z}ij}\ z \bar{z}+O(x^3) \,.
\end{eqnarray}
Applying formulas (\ref{dA}, \ref{dAdA}), we derive HEE of the total derivative terms. We list the results below.

For $\Box R$, we get
\begin{eqnarray}\label{ddRDong}
S_{\text{G-Wald}}&=&4\pi \int d^dy \sqrt{g} \big{[} (\text{Tr} K)^2-3 \text{Tr}K^2 +2 \text{Tr}Q^{\ a}_{0\ a}-24T_0\big{]} \,,\nonumber\\
S_{\text{Anomaly}}&=&-4\pi \int d^dy \sqrt{g} \big{[} (\text{Tr} K)^2-3 \text{Tr}K^2 +2 \text{Tr}Q^{\ a}_{0\ a}-24T_0\big{]} \,,\nonumber\\
S_{\text{HEE}}&=&0 \,.
\end{eqnarray}

For  $\Box R^2$, we have
\begin{eqnarray}\label{ddRR}
S_{\text{G-Wald}}&=&8\pi \int d^dy \sqrt{g} R \big{[} (\text{Tr} K)^2-3 \text{Tr}K^2 +2 \text{Tr}Q^{\ a}_{0\ a}-24T_0\big{]} \,,\nonumber\\
S_{\text{Anomaly}}&=&-8\pi \int d^dy \sqrt{g} R \big{[} (\text{Tr} K)^2-3 \text{Tr}K^2 +2 \text{Tr}Q^{\ a}_{0\ a}-24T_0\big{]} \,,\nonumber\\
S_{\text{HEE}}&=&0 \,.
\end{eqnarray}

The calculations for $\Box (R_{\mu\nu}R^{\mu\nu})$ are quite complicated. For simplicity, we work in 3d bulk spacetime and obtain
\begin{eqnarray}\label{ddR2}
S_{\text{G-Wald}}&=&32 \pi \int dy \sqrt{g} [\frac{5}{2} \left(K_{\bar{z}}^2 Q_{zz}+K_z^2 Q_{\bar{z}\bar{z}}\right)-9K_{\bar{z}}^2K_z^2+6 K_{\bar{z}}K_z (Q_{0\ z\bar{z}}+2 Q_{1\ z\bar{z}}-4 T_0-2 T_1) \nonumber\\
&&\ \ \ \ -2Q_{\bar{z}\bar{z}} Q_{zz} -6 \left(Q_{0\ z\bar{z}} (Q_{1\ z\bar{z}}-2 T_0)+Q_{1\ z\bar{z}}^2-2 Q_{1\ z\bar{z}}(2 T_0+T_1)+12 T_0 (T_0+T_1)\right)] \,,\nonumber\\
S_{\text{Anomaly}}&=&-32 \pi \int dy \sqrt{g} [\frac{5}{2} \left(K_{\bar{z}}^2 Q_{zz}+K_z^2 Q_{\bar{z}\bar{z}}\right)-9K_{\bar{z}}^2K_z^2+6 K_{\bar{z}}K_z (Q_{0\ z\bar{z}}+2 Q_{1\ z\bar{z}}-4 T_0-2 T_1) \nonumber\\
&&\ \ \ \  -2Q_{\bar{z}\bar{z}} Q_{zz} -6 \left(Q_{0\ z\bar{z}} (Q_{1\ z\bar{z}}-2 T_0)+Q_{1\ z\bar{z}}^2-2 Q_{1\ z\bar{z}}(2 T_0+T_1)+12 T_0 (T_0+T_1)\right)] \,,\nonumber\\
S_{\text{HEE}}&=&0 \,.
\end{eqnarray}
Similarly, for $\Box (R_{\mu\nu\rho\sigma}R^{\mu\nu\rho\sigma})$ in 3d spacetime we find
\begin{eqnarray}\label{ddR4}
S_{\text{G-Wald}}&=&128 \pi  \int dy \sqrt{g}[\frac{5}{2} \left(K_{\bar{z}}^2 Q_{zz}+K_z^2 Q_{\bar{z}\bar{z}}\right)-5 K_{\bar{z}}^2 K_z^2+2 K_{\bar{z}}K_z (Q_{0\ z\bar{z}}+2 Q_{1\ z\bar{z}})\nonumber\\
&&\ \ \ \ \ \ \ \ \ \ \ \ -2 (Q_{\bar{z}\bar{z}}Q_{zz}+Q_{1\ z\bar{z}}(Q_{0\ z\bar{z}}+Q_{1\ z\bar{z}})+18 T_0(T_0+T_1))] \,,\nonumber\\
S_{\text{Anomaly}}&=&-128 \pi  \int dy \sqrt{g}[\frac{5}{2} \left(K_{\bar{z}}^2 Q_{zz}+K_z^2 Q_{\bar{z}\bar{z}}\right)-5 K_{\bar{z}}^2 K_z^2+2 K_{\bar{z}}K_z (Q_{0\ z\bar{z}}+2 Q_{1\ z\bar{z}})\nonumber\\
&&\ \ \ \ \ \ \ \ \ \ \ \ -2 (Q_{\bar{z}\bar{z}}Q_{zz}+Q_{1\ z\bar{z}}(Q_{0\ z\bar{z}}+Q_{1\ z\bar{z}})+18 T_0(T_0+T_1))] \,,\nonumber\\
S_{\text{HEE}}&=&0 \,.
\end{eqnarray}
Remarkably, the above results show that the generalized Wald entropy and the anomaly-like entropy always exactly cancel for total derivative actions.

Before the end of this section, we provide some details of the calculations for HEE of $\Box R$. Focusing on the linear terms of $A$ which are relevant to the generalized Wald entropy, we get
\begin{eqnarray}\label{boxRWald}
\sqrt{G}\Box R=e^{-2A}\sqrt{g}\big{[} && 4 ((\text{Tr} K)^2-3 \text{Tr}K^2 +2 \text{Tr}Q^{\ a}_{0\ a}-16T_0+64e^{2A}T_1) \partial_{z}\partial_{\bar{z}}A\nonumber\\
&-&2((\text{Tr} K)^2
-2 \text{Tr}K^2 + \text{Tr}Q^{\ a}_{0\ a}-32T_0-112e^{2A}T_1)(z \partial_{z}^2\partial_{\bar{z}}A+\bar{z} \partial_{z}\partial_{\bar{z}}^2A )\nonumber\\
&+&(32T_0+48e^{2A}T_1)(z^2 \partial_{z}^3\partial_{\bar{z}}A+\bar{z}^2 \partial_{z}\partial_{\bar{z}}^3A )\nonumber\\
&-&4((\text{Tr} K)^2
-2 \text{Tr}K^2 + \text{Tr}Q^{\ a}_{0\ a}+8T_0)( z\bar{z} \partial_{z}^2\partial_{\bar{z}}^2A )\big{]}+\cdots \,.
\end{eqnarray}
Applying eqs.~(\ref{dA}), we obtain the first formula in eq.~(\ref{ddRDong}):
\begin{eqnarray}\label{coneddRWald}
S_{\text{G-Wald}}&=&4\pi \int d^dy \sqrt{g} \big{[} (\text{Tr} K)^2-3 \text{Tr}K^2 +2 \text{Tr}Q^{\ a}_{0\ a}-24T_0\big{]} \,.
\end{eqnarray}
Note that the generalized Wald entropy is nonzero for $\Box R$. 

Let us proceed to compute the anomaly-like entropy from $\Box R$. Focusing on the relevant terms eq.~(\ref{dAdA}), we have
\begin{eqnarray}\label{boxRAnomaly}
\sqrt{G}\Box R=&-&8\sqrt{g}  e^{-2A}[(\text{Tr} K)^2-3 \text{Tr}K^2 +2 \text{Tr}Q^{\ a}_{0\ a}-8T_0-8e^{2A}T_1) \partial_zA \partial_{\bar{z}}A\nonumber\\
&+&64(-2T_0+e^{2A}T_1) (z\partial_z^2A \partial_{\bar{z}}A+c.c)\nonumber\\
&-&32T_0 (z^2\partial_z^3A \partial_{\bar{z}}A+c.c)\nonumber\\
&+&64e^{2A}T_1 (z\bar{z}\partial_z^2A \partial_{\bar{z}}^2A)]+\cdots \,.\nonumber\\
\end{eqnarray}
Applying eq.~(\ref{dAdA}), we derive the second formula in eq.~(\ref{ddRDong}):
\begin{eqnarray}\label{coneddRAnomaly}
S_{\text{Anomaly}}&=&-4\pi \int d^dy \sqrt{g} \big{[} (\text{Tr} K)^2-3 \text{Tr}K^2 +2 \text{Tr}Q^{\ a}_{0\ a}-24T_0\big{]}\,.
\end{eqnarray}
As expected, the generalized Wald entropy (\ref{coneddRWald}) and the anomaly-like entropy (\ref{coneddRAnomaly}) cancel. 

To summarize, by applying the methods of \cite{Dong, Miao} we find that the entropy from covariant total derivative terms in the gravitational action is indeed zero.

\section{Comparison with Astaneh-Patrushev-Solodukhin}\la{sec:compare}

In this section, we compare our results with those obtained in recent papers by Astaneh, Patrushev and Solodukhin (APS) \cite{Astaneh,Astaneh2}.  For simplicity we will illustrate the differences by using the $\Box R$ example.  The works \cite{Astaneh,Astaneh2} use the prescription for regularization developed in \cite{Solodukhin}, which is quite different from the Lewkowycz-Maldacena prescription \cite{Maldacena}. The main differences are that the regularized cone used in \cite{Astaneh,Astaneh2} does not approach the singular cone away from the conical singularity, and that they do not subtract off the on-shell action of the singular cone. As a result, they sometimes get a nonzero entropy from total derivative terms. 

Let us briefly review the prescription used in \cite{Astaneh,Astaneh2}. They propose to write the regularized conical metric as\footnote{Note that the cone here has a conical excess of $2\pi(n-1)$.  For integer $n$ it can be constructed by gluing $n$ copies of the $B_1$ bulk geometry.  This is different from our orbifold picture in the previous sections in which the cone has a conical deficit of $2\pi(1-\fr{1}{n})$.}
\begin{eqnarray}\label{conicalmetric}
ds^2=f_n(r)dr^2+r^2d\tau^2+[g_{ij}+2K^a_{ij} n^a r^n +K^a_{im}K^{b m}_{\ j}n^an^b r^{2n}+\cdots ]dy^idy^j \,,
\end{eqnarray}
where $f_n=\frac{r^2+b^2 n^2}{r^2+b^2}$, $n^1=\cos \tau$, $n^2=\sin \tau$, and $\tau\sim\tau+2n\pi$. Note that we have $f_n\to n^2$ for $r\to 0$, ensuring that there is no conical singularity when we identify $\tau$ with $\tau+2\pi n$. Using the above regularized metric, they derive the generalized gravitational entropy as
\begin{eqnarray}\label{entropyFPS}
&&S_{\text{GGE}}=\lim_{n\to 1}(n\partial_n -1)I_{\rm reg} \,,
\end{eqnarray}
with $I_{\rm reg}$ the gravitational action of the regularized cone. 

Before proceeding, let us point out two differences between the APS prescription and the Lewkowycz-Maldacena prescription used in the previous sections.  Even though both methods use regularized cones, an important difference is that at large $r$ the metric \er{conicalmetric} does not approach the singular conical metric
\be\la{scone}
ds^2=dr^2+r^2d\tau^2+[g_{ij}+2K^a_{ij} n^a r +\cdots ]dy^idy^j \,.
\ee
The regularizing procedure of replacing $r$ by $r^n$ in the extrinsic curvature term was first proposed in \cite{Solodukhin}, but it is not a local modification of the cone near the conical singularity.  The second important difference is that unlike the Lewkowycz-Maldacena prescription \er{eeir} or \er{HEE}, the on-shell action of the singular cone is not subtracted off in \er{entropyFPS}.

Now we are ready to reproduce the calculation of the entropy from $\Box R$ by using the APS approach. By dimensional analysis, we note that the entropy is of order $O(K^2)$. Focusing on this order, we obtain 
\begin{eqnarray}\label{FPSBoxR}
&&\int_{0}^{2\pi n}d\tau\int_{0}^{r_0}dr\sqrt{G}\Box R=\int_{0}^{2\pi n}d\tau \sqrt{G}G^{rr}\partial_r R |^{r= r_0}_{r=0}\nonumber\\
&&=4\pi (n-1) r^{2(n-1)}\frac{r^6 \Tr K^2+c_1 b^2r^4+c_2b^4r^2+((\Tr K)^2-\Tr K^2)b^6}{(b^2+r^2)^3}|_{r=0}^{r=r_0}+O(n-1)^2\nonumber\\
&&=4\pi (n-1) \Tr K^2+O(n-1)^2
\end{eqnarray}
where we have used the fact that $r_0\gg b$ in the above derivation. The exact expressions of $c_1, c_2$ are irrelevant for the calculation. Remarkably, only the terms at $r=r_0$ contribute to the final result, while the terms at $r=0$ vanish because $n>1$.

Using \er{entropyFPS} with this result we would be tempted to conclude that the total derivative action $\Box R$ contributes to the generalized gravitational entropy \cite{Astaneh,Astaneh2}:
\begin{eqnarray}\label{FPSBoxR1}
S_{\text{GGE,APS}}&=& 4\pi \int d^dy\sqrt{g}\  \text{Tr} K^2 \,.
\end{eqnarray}
However, if we choose a regularization of the cone such that it approaches the singular cone \er{scone} away from the conical singularity, and subtract off the on-shell action of the singular cone in \er{entropyFPS}, we would find that the entropy from $\Box R$ is zero as in Sec.~\re{sec:boxr}.

\section{Which prescription is correct?}\la{sec:problems}

As we saw in the previous section, the Lewkowycz-Maldacena prescription and the APS prescription generally give different results for the entropy.  While a lot of confidence is usually given to the Lewkowycz-Maldacena prescription because of the underlying argument reviewed in Sec.~\re{sec:lm}, in this section we would like to be more open-minded and ask which prescription is correct.
We find that the holographic and the field-theoretic universal terms of the entanglement entropy do not match if total derivative terms produce a nonzero entropy. Furthermore, the second law of black hole thermodynamics could be violated if the entropy of total derivative terms is nonzero. Thus it only seems reasonable if total derivative terms in the action do not contribute to the entropy. 

\subsection{Entropy discrepancy}

In this section, we show that there is entropy discrepancy between the holographic and the field-theoretic results by using the APS prescription. For simplicity, we focus on the case of a 4-dimensional field theory. A discussion for 6-dimensional field theories is in \cite{Miao1}.

Let us start with the following bulk action in a 5-dimensional spacetime
\begin{eqnarray}\label{action1}
S&=& \frac{1}{16\pi} \int d^5x \sqrt{G} [R-2\Lambda + \beta \Box R]
\end{eqnarray}
where $\Lambda=-\frac{6}{l^2}$ is the cosmological constant and $\beta$ is a free parameter. 

By applying the APS prescription \cite{Astaneh, Astaneh2}, we obtain the holographic entanglement entropy for action (\ref{action1}) as
\begin{eqnarray}\label{HEEaction1}
S_{\text{HEE}}&=& \frac{1}{4} \int d^3y \sqrt{g} [1-\beta \text{Tr} K^2]\,. 
\end{eqnarray}
Note that we work in the Lorentzian signature in this section, which differs from its Euclidean
form (\ref{FPSBoxR1}) by a minus sign. By applying the method of \cite{Theisen}, it is not difficult to derive the universal terms of the entanglement entropy as
\begin{eqnarray}\label{holographiclog}
S_{\Sigma}|_{\log}=\log(\ell/\delta)\frac{1}{2\pi} \int_{\Sigma} \big{[} c (C^{ijkl}h_{ik}h_{jk}-\Tr \bar{k}^2)-a R_{\Sigma}-\frac{\pi}{2}\beta \Tr \bar{k}^2\big{]}\,,
\end{eqnarray}
where $C_{ijkl}$ is the Weyl tensor and $\bar{k}$ is the traceless part of the extrinsic curvature on the entangling surface $\Sigma$. The central charges $a$ and $c$ are given in Planck units by 
\begin{eqnarray}\label{centralcharge}
a=\frac{\pi l^3}{8} \,,\quad c=\frac{\pi l^3}{8} \,. 
\end{eqnarray} 
Note that eq.~(\ref{holographiclog}) is conformally invariant. 

Following the approach of \cite{Henningson, Miao3}, one can derive the holographic Weyl anomaly for action (\ref{action1}). An advantage of the approach of \cite{Miao3} is that one does not need to solve the equation of motion in 5-dimensional (or 7-dimensional) bulk theories. We obtain 
\begin{eqnarray}\label{Weylanomalyaction1}
\<T^i_{\ i}\>=\frac{c}{16\pi^2}C_{ijkl}C^{ijkl}-\frac{a}{16\pi^2}E_4 \,.
\end{eqnarray}
Remarkably, the total derivative term $\Box R$ does not contribute to the holographic Weyl anomaly. In the field-theoretic approach, we can derive the universal terms of the entanglement entropy as the `entropy' of the Weyl anomaly \cite{Solodukhin:2008dh,Hung,Myers}. We get
\begin{eqnarray}\label{fieldlog}
S'_{\Sigma}|_{\log}=\log(\ell/\delta)\frac{1}{2\pi} \int_{\Sigma} \big{[} c (C^{ijkl}h_{ik}h_{jk}-\Tr \bar{k}^2)-a R_{\Sigma}\big{]}\,.
\end{eqnarray}
Clearly, the holographic result eq.~(\ref{holographiclog}) and the field-theoretic result eq.~(\ref{fieldlog}) do not match, unless the entropy from total derivative terms vanishes. In general, a total derivative term may appear in the Weyl anomaly
\begin{eqnarray}\label{Anyanomalyaction1}
\<T^i_{\ i}\>=\frac{c}{16\pi^2}C_{ijkl}C^{ijkl}-\frac{a}{16\pi^2}E_4+ \frac{\lambda}{16\pi^2} D^iD_i \bar{R} \,,
\end{eqnarray}
where $D_i$ and $\bar{R}$ are the covariant derivative and the Ricci scalar on the boundary. Here $\lambda$ is a parameter that depends on the regularization. We have $\lambda=0$ for the holographic Weyl anomaly (\ref{Weylanomalyaction1}).  We choose the holographic Weyl anomaly in the field-theoretic calculations. This is because in deriving the holographic result (\ref{holographiclog}) we have used a holographic regularization with the cutoff surface $z=\delta$ where $z$ is the radial coordinate of AdS. Thus it is natural to use the same regularization for the Weyl anomaly. As a result, we get the holographic Weyl anomaly with $\lambda=0$. Note that even if we choose a different regularization for the Weyl anomaly, the holographic entropy and the field-theoretic entropy still do not match. From the general Weyl anomaly (\ref{Anyanomalyaction1}), we can derive the universal terms of entanglement entropy as
\begin{eqnarray}\label{fieldlog2}
S''_{\Sigma}|_{\log}=\log(\ell/\delta)\frac{1}{2\pi} \int_{\Sigma} \big{[} c (C^{ijkl}h_{ik}h_{jk}-\Tr \bar{k}^2)-a R_{\Sigma}-\frac{\lambda}{2}\Tr k^2 \big{]} \,.
\end{eqnarray}
Note that the above equation is not conformally invariant for nonzero $\lambda$, while the holographic universal term (\ref{holographiclog}) is conformally invariant. Therefore, the holographic result (\ref{holographiclog}) and the field-theoretic result (\ref{fieldlog}, \ref{fieldlog2}) cannot match, unless the entropy from total derivatives vanishes.

\subsection{Violation of the universality of corner entanglement}

In this section, we show that the results of APS \cite{Astaneh,Astaneh2} do not agree with the conjecture of \cite{Bueno:2015rda} for the universal part of the corner contribution to entanglement entropy.\footnote{This conjecture was recently proven in \cite{Faulkner:2015csl}.} Since the conjecture of~\cite{Bueno:2015rda} has passed several quite general tests, it suggests that the entropy from total derivative terms should vanish.  

Let us first briefly review the works of \cite{Bueno:2015rda,Bueno:2015xda}.
The entanglement entropy (EE) of some region $V$ in 3d CFTs takes the form
\begin{eqnarray}\label{3dEE}
S=B\ H/\delta -a(\Omega) \log(H/\delta)+ O(1) \,,
\end{eqnarray}
where $\delta$ is a short-distance cutoff, $B$ is a constant, and $H$ denotes the size of the entangling surface. The first term of eq.~(\ref{3dEE}) is the `area law' contribution to EE and the second logarithmic term appears only if the entangling surface has a sharp corner. For pure state, we have $a(\Omega)=a(2\pi-\Omega)$ due to the fact $S(V)=S(\bar{V})$. Thus we have
\begin{eqnarray}\label{acondition1}
a(\Omega\to \pi)\simeq \sigma (\pi-\Omega)^2
\end{eqnarray}
in the smooth limit. Recently, it is conjectured that
\begin{eqnarray}\label{conjecture}
\sigma/C_T=\pi^2/24
\end{eqnarray}
is a universal relation for all CFTs in three dimensions \cite{Bueno:2015rda}. Here $C_T$ is the central charge appearing in the stress tensor correlator
\begin{eqnarray}\label{twopoint}
\<T_{\mu\nu}(x)T_{\lambda\rho}(0)\>=\frac{C_T}{|x|^{2d}}I_{\mu\nu,\lambda\rho}(x)
\end{eqnarray}
with $I_{\mu\nu,\lambda\rho}$ a dimensionless tensor fixed by symmetry. 

This conjecture was tested in \cite{Bueno:2015rda,Bueno:2015xda} by studying some holographic models, free scalars, and free fermions. It was later proved for CFTs dual to general higher curvature gravity \cite{Miao:2015dua,Bueno:2015lza}.
For simplicity, below we take Einstein gravity and curvature-squared gravity as examples to illustrate the universality of corner entanglement. 

Consider the following action
\begin{eqnarray}\label{HDG0}
I=\frac{1}{16\pi G}\int d^{4}x\sqrt{g}\big[R+\frac{6}{l^2} + \lambda l^2  \bar{R}_{\mu\nu\sigma\rho}\bar{R}^{\mu\nu\sigma\rho} \big]
\end{eqnarray}
where $\bar{R}_{\mu\nu\sigma\rho}=R_{\mu\nu\sigma\rho}+\frac{1}{l^2}(g_{\mu\sigma}g_{\nu\rho}-g_{\mu\rho}g_{\nu\sigma})$. For simplicity, we focus on $AdS_4$ where $\bar{R}_{\mu\nu\sigma\rho}=0$.  Following \cite{Solodukhin,Dong,Camps}, we get the holographic entanglement entropy of the model (\ref{HDG0}) as 
\begin{eqnarray}\label{CornerHEE0}
S&=&\frac{1}{4G}\int d^2y\sqrt{h} (1-2\lambda l^2 \text{Tr} K^2 ) \,.
\end{eqnarray}
Using the Gauss-Codazzi equations in $AdS_4$, we can rewrite
 \begin{eqnarray}\label{trKK}
 \int_{\Sigma} d^{2}y \sqrt{h}\text{Tr} K^2=\int_{\Sigma} d^{2}y \sqrt{h} \[-\frac{2}{l^2} -\mathcal{R}+(\text{Tr} K)^2 \] \simeq -\frac{2}{l^2}\int_{\Sigma} d^{2}y \sqrt{h}
 \end{eqnarray}
where in the last step we have ignored  $(\text{Tr} K)^2$ and the total derivative term $\sqrt{h}\mathcal{R}$ as they do not contribute to the universal part of corner entanglement\footnote{According to \cite{Bueno:2015rda,Bueno:2015xda}, $(\text{Tr} K)^2$ give higher order terms near the minimal surface and thus can be ignored. \cite{Bueno:2015rda,Bueno:2015xda} also show that $\mathcal{R}$ does not contribute to the universal term $a(\Omega)$.}.  We can therefore rewrite eq.~(\ref{CornerHEE0}) as
 \begin{eqnarray}\label{HEE0}
 S&=&(1+4\lambda)\int d^2y\frac{\sqrt{h} }{4G}
 \end{eqnarray}
 which is equivalent to the entropy of Einstein gravity up to an overall factor. 
As a result, we have $a(\Omega)=(1+4\lambda) a_E(\Omega)$ and thus $\sigma=(1+4\lambda)\sigma_E$, where $E$ denotes Einstein gravity.

Now let us discuss the central charge $C_T$ appearing in eq.~(\ref{twopoint}). A standard holographic calculation of $C_T$ for Einstein gravity gives
\begin{eqnarray}\label{CTE0}
C_{T,E}=\frac{3\ l^2}{\pi^3G} \,.
\end{eqnarray}
The situation is a little more complicated for higher curvature gravity. That is because, in addition to the usual massless spin-two graviton, massive modes and ghost modes with $M^2 \sim 1/(\lambda l^2) $ also appear in higher curvature gravity. To suppress these modes, it is natural to work in the perturbative framework with $\lambda \ll 1$. Consider the metric fluctuations in the $AdS_4$ background with the gauge conditions $\bar{\nabla}^{\mu}h_{\mu\nu}=0$ and $g^{\mu\nu}h_{\mu\nu}=0$, we can derive the linearized Einstein equations as
\begin{eqnarray}\label{EOME}
-\frac{1}{2}[\bar{\Box}+\frac{2}{\bar{L}^2}]h_{\mu\nu}=8\pi G T_{\mu\nu} \,.
\end{eqnarray}
Similarly, we can derive the linearized equation of motion for the holographic model  (\ref{HDG0}) as
\begin{eqnarray}\label{EOMHDG0}
-\frac{1+4\lambda}{2}[\bar{\Box}+\frac{2}{\bar{L}^2}]h_{\mu\nu}-2\lambda[\bar{\Box}+\frac{2}{\bar{L}^2}]^2h_{\mu\nu}=8\pi G T_{\mu\nu} \,.
\end{eqnarray}
Clearly, the second term of the above equation is suppressed near the physical pole, i.e. $[\bar{\Box}+\frac{2}{\bar{L}^2}]h_{\mu\nu}\sim 0$. Comparing eq.~(\ref{EOMHDG0}) with eq.~(\ref{EOME}), we notice that the effective Newton constant of the holographic model (\ref{HDG0}) is $G_{eff}=G/(1+4\lambda)$. From eq.~(\ref{CTE0}), we get $C_T=(1+4\lambda) C_{T,E}$. Recall that we have $\sigma=(1+4\lambda) \sigma_E$ from eq.~(\ref{HEE0}), and we finally obtain
\begin{eqnarray}\label{Myersresult}
\frac{\sigma}{C_T}=\frac{\sigma_E}{C_{T,E}}
\end{eqnarray}
which agrees with the conjecture (\ref{conjecture}).

Now let us discuss the effects of total derivative terms in the gravitational action. For simplicity, we focus on the following action
\begin{eqnarray}\label{HDG0TD}
I=\frac{1}{16\pi G}\int d^{4}x\sqrt{g}\big[R+\frac{6}{l^2} + \lambda  \bar{R}_{\mu\nu\sigma\rho}\bar{R}^{\mu\nu\sigma\rho} + \beta \Box R \big].
\end{eqnarray}
Using eqs.~(\ref{FPSBoxR1}, \ref{CornerHEE0}, \ref{trKK}), we obtain the entropy for the holographic model (\ref{HDG0TD}) as
\begin{eqnarray}\label{HEE0TD}
S&=&\frac{1}{4G}\int d^2y\sqrt{h} \big(1-(2\lambda+\beta) \text{Tr} K^2 \big)\nonumber\\
&\simeq&(1+4\lambda+2\beta)\int d^2y\frac{\sqrt{h}}{4G} \,,
\end{eqnarray}
which yields $\sigma=(1+4\lambda+2\beta)\sigma_E$. As for the central charge $C_T$, since total derivatives do not contribute to the equation of motion, using eq.~(\ref{EOMHDG0}) we get $C_T=(1+4\lambda)C_{T,E}$. Now it is clear that the conjecture of \cite{Bueno:2015rda} is violated 
\begin{eqnarray}\label{MyersresultTD}
\frac{\sigma}{C_T}=\frac{1+4\lambda+2\beta}{1+4\lambda}\frac{\sigma_E}{C_{T,E}}\neq\frac{\sigma_E}{C_{T,E}} \,,
\end{eqnarray}
unless the entropy from total derivative terms vanishes.

\subsection{Violation of the second law}

In this section, we prove that the second law of black hole thermodynamics can be violated if the entropy from total derivatives is nonzero. For simplicity, we focus on linearized metric perturbations on stationary black holes with a regular bifurcation surface. It is found that the linearized second law is obeyed by f(Lovelock) gravity \cite{Wall1}, curvature-squared gravity \cite{Wall2}, and higher derivative gravity~\cite{Wall3}\footnote{See also \cite{Bhattacharjee:2015qaa} for discussions beyond the linearized second law.}. However, if total derivatives produce nonzero entropy, the linearized second law can be violated as we shall show below. To obey the second law, the entropy from total derivatives must therefore vanish.

Consider the Einstein-Hilbert action plus a total derivative term and a matter action
\begin{eqnarray}\label{action2}
S&=& \frac{1}{16\pi} \int d^Dx \sqrt{G} [R + \nabla_{\mu} J^{\mu}]+S_{\text{M}} \,.
\end{eqnarray}
It is well-known that total derivatives do not affect the equation of motion. Thus we have
\begin{eqnarray}\label{equations2}
R_{\mu\nu}-\frac{R}{2}G_{\mu\nu}=8\pi T_{\mu\nu} \,.
\end{eqnarray}
By using the APS prescription, the entropy from the total derivative term $\nabla_{\mu} J^{\mu}$ is nonzero. Let us denote the entropy density of the higher derivative correction (which in the case above is $\nabla_{\mu} J^{\mu}$) by $4\pi \rho$. The total entropy becomes
\begin{eqnarray}\label{entropy2}
S_{\rm GGE}=\frac{1}{4}\int d^3y\sqrt{g}\  (1+\rho) \,.
\end{eqnarray}
Define the change of entropy per unit area as
\begin{eqnarray}\label{chargeentropy2}
\Theta=\frac{d\rho}{dt}+\theta_k(1+\rho)
\end{eqnarray}
where $\frac{d}{dt}=k^{\mu}\nabla_{\mu}$, $\theta_k$ is the expansion, and $k^{\mu}$ is the null generator on the horizon.  Following \cite{Wall2} and neglecting some higher order terms in the Raychaudhuri equation, we obtain the evolution equation of $\Theta$:
\begin{eqnarray}\label{evolution2}
\frac{d \Theta}{dt}-\kappa \Theta=-8\pi T_{kk}+\nabla_k\nabla_k\rho-\rho R_{kk}+H_{kk}
\end{eqnarray}
where $\kappa$ is the surface gravity, and $H_{kk}$ is the contribution to the equation of motion from higher curvature terms (which is zero in the case of a total derivative).  It turns out that for general higher curvature gravity, $(\nabla_k\nabla_k\rho-\rho R_{kk}+H_{kk})$ vanishes at the linearized order \cite{Wall1,Wall2,Wall3}. As a result, the linearized second law is obeyed. 

Let us briefly review the argument of \cite{Wall2}. Consider a black hole that begins and ends in a stationary state, but at some intermediate time one perturbs it with a stress tensor $T_{\mu\nu}$ that obeys the null energy condition $T_{kk}\ge 0$. Recall that we have $(\nabla_k\nabla_k\rho-\rho R_{kk}+H_{kk})=O(\epsilon^2)$ for general higher curvature gravity \cite{Wall1,Wall2,Wall3}. At the linearized order, we obtain
\begin{eqnarray}\label{evolutionWall}
\frac{d \Theta}{dt}-\kappa \Theta=-8\pi T_{kk} \le 0 \,.
\end{eqnarray}
If $\Theta <0$ at some moment, we have $\frac{d \Theta}{dt}<0$ due to $\kappa>0$, and therefore $\Theta$ would never be zero in future. Thus we must always have $\Theta \ge 0$ and the linearized second law is obeyed. 

Now let us return to our case with total derivative terms. Recall that total derivatives do not affect the equation of motion $H_{kk}=0$. If they contribute to the entropy, i.e.\ $\rho \neq 0$, $(\nabla_k\nabla_k\rho-\rho R_{kk}+H_{kk})$ would generally be nonzero. As a result, the above argument breaks down and the linearized second law may be violated. Below we give an example where this indeed happens. Now let us focus on $\nabla_{\mu}J^{\mu}=\beta \Box R$. From eq.~(\ref{FPSBoxR1}) we get $\rho=-\beta \text{Tr} K^2$. Note that we work in Lorentzian signature in this section. Let us take the Vaidya metric as an example
\begin{eqnarray}\label{Vaidyametric}
ds^2=-\(1-\frac{2M(v)}{r}\)dv^2+2dv dr +r^2 d\Sigma^2_{D-2} \,.
\end{eqnarray}
The energy density is $\frac{M'(v)}{4\pi r^2}>0$, and the expansion is given by $\theta_k=\frac{r-2M(v)}{r^2}\ge 0$.
 
After some calculations, we derive 
\begin{eqnarray}\label{argument1}
\Theta=\frac{d\rho}{dt}+\theta_k(1+\rho)=\frac{(r-2 M(v)) \left(-2 \beta  M(v)+r^3\right)+4 \beta  r^2 M'(v)}{r^5} \,.
\end{eqnarray}
According to \cite{Wall2}, the location of event horizon $r=r(v)$ can be obtained by solving the equation
\begin{eqnarray}\label{horizon}
r'=\frac{1-\frac{2M(v)}{r}}{2},
\end{eqnarray}
where $r'=\frac{dr(v)}{dv}$. 
Using eq.~(\ref{horizon}) to rewrite eq.~(\ref{argument1}), we obtain
\begin{eqnarray}\label{argument2}
\Theta=\frac{2r r'-4\beta r''}{r^2} \,.
\end{eqnarray}
From the positivity of the expansion $\theta_k=\frac{r-2M(v)}{r^2}$ and the energy density $\frac{M'(v)}{4\pi r^2}$, we get two constraints for $r(v)$
\begin{eqnarray}\label{horizoncondition}
&&r'\ge 0 \,, \\
&&r'-2r'^2-2r r''\ge 0 \,.\label{horizoncondition1}
\end{eqnarray}
If we require $M>0$, we have one additional constraint $r'<\frac{1}{2}$. 

For a fixed parameter $\beta$, by choosing suitable evolution of the Vaidya metric
\begin{eqnarray}\label{argument3}
&&0<\frac{rr'}{2r''}(v_0)<\beta \,, \ \text{when} \ \beta>0\\
&&0>\frac{rr'}{2r''}(v_0)>\beta \,, \ \text{when} \ \beta<0\label{argument4}
\end{eqnarray}
at some moment $v=v_0$, we can always make $\Theta(v_0)<0$. To demonstrate this explicitly, we study the following toy model
\begin{eqnarray}\label{example1}
r(v)=\frac{\sqrt{|\beta|}}{2}\bigg(\ 1+\frac{1}{2}\tanh\frac{v}{2\sqrt{|\beta|}} \ \bigg) \,,
\end{eqnarray}
which satisfies the constraints (\ref{horizoncondition}, \ref{horizoncondition1}) and $M>0$ for $-\infty<v<\infty$. One can check that the second law is violated, i.e., $\Theta(v)<0$, in the above toy model for $v< -2\sqrt{|\beta|} \tanh ^{-1}\left(\frac{2}{9}\right)$ when $\beta>0$, and for $v> 2\sqrt{|\beta|} \tanh ^{-1}\left(\frac{2}{7}\right)$ when $\beta<0$. In conclusion, the second law of black hole thermodynamics can be violated unless the entropy from total derivative terms vanishes. 

\section{Conclusion}\la{sec:conclude}

By applying the Lewkowycz-Maldacena method, we have investigated the generalized gravitational entropy from total derivative terms in the gravitational action. In contrast to \cite{Astaneh,Astaneh2}, we find that the entropy from total derivative terms vanishes. The Lewkowycz-Maldacena prescription and the APS prescription \cite{Astaneh,Astaneh2} generally give different results for the entropy.  We find that the APS prescription would lead to the conclusion that the holographic entropy and the field-theoretic entropy do not match. Furthermore, the second law of black hole thermodynamics could be violated if the entropy from total derivative terms is nonzero. These results give us more confidence that the generalized gravitational entropy from total derivative terms in the action vanishes.

\section*{Acknowledgements}

We thank Joan Camps, Ethan Dyer, Sean Hartnoll, Ling-Yan Hung, Kristan Jensen, Sergey Solodukhin, Tadashi Takayanagi, Stefan Theisen, and Aron Wall for valuable comments and discussions. XD is supported by the National Science Foundation under grant PHY-1314311 and by a Zurich Financial Services Membership at the Institute for Advanced Study. RXM is supported by Sino-German (CSC-DAAD) Postdoc Scholarship Program.


\begin{thebibliography}{00}

\bibitem{Bekenstein:1973ur} 
  J.~D.~Bekenstein,
  ``Black holes and entropy,''
  Phys.\ Rev.\ D {\bf 7}, 2333 (1973).

\bibitem{Bardeen:1973gs} 
  J.~M.~Bardeen, B.~Carter and S.~W.~Hawking,
  ``The Four laws of black hole mechanics,''
  Commun.\ Math.\ Phys.\  {\bf 31}, 161 (1973).

\bibitem{Hawking:1974sw} 
  S.~W.~Hawking,
  ``Particle Creation by Black Holes,''
  Commun.\ Math.\ Phys.\  {\bf 43}, 199 (1975)
  [Commun.\ Math.\ Phys.\  {\bf 46}, 206 (1976)].
  
\bibitem{GH}
  G.~W.~Gibbons and S.~W.~Hawking,
  ``Action Integrals and Partition Functions in Quantum Gravity,''
  Phys.\ Rev.\ D {\bf 15}, 2752 (1977).
    
\bibitem{RT}
  S.~Ryu and T.~Takayanagi,
  ``Holographic derivation of entanglement entropy from AdS/CFT,''
  Phys.\ Rev.\ Lett.\  {\bf 96}, 181602 (2006)
  [hep-th/0603001].

\bibitem{Maldacena}
  A.~Lewkowycz and J.~Maldacena,
  ``Generalized gravitational entropy,''
  JHEP {\bf 1308}, 090 (2013)
  [arXiv:1304.4926 [hep-th]].

\bibitem{Wald}
  R.~M.~Wald,
  ``Black hole entropy is the Noether charge,''
  Phys.\ Rev.\ D {\bf 48}, 3427 (1993)
  [gr-qc/9307038].

\bibitem{Solodukhin}
  D.~V.~Fursaev, A.~Patrushev and S.~N.~Solodukhin,
  ``Distributional Geometry of Squashed Cones,''
  Phys.\ Rev.\ D {\bf 88}, no. 4, 044054 (2013)
  [arXiv:1306.4000 [hep-th]].

\bibitem{Dong}
  X.~Dong,
  ``Holographic Entanglement Entropy for General Higher Derivative Gravity,''
  JHEP {\bf 1401}, 044 (2014)
  [arXiv:1310.5713 [hep-th]].

\bibitem{Camps}
  J.~Camps,
  ``Generalized entropy and higher derivative Gravity,''
  JHEP {\bf 1403}, 070 (2014)
  [arXiv:1310.6659 [hep-th]].

\bibitem{Astaneh}
  A.~F.~Astaneh, A.~Patrushev and S.~N.~Solodukhin,
  ``Entropy vs Gravitational Action: Do Total Derivatives Matter?,''
  arXiv:1411.0926 [hep-th].

\bibitem{Astaneh2}
  A.~F.~Astaneh, A.~Patrushev and S.~N.~Solodukhin,
  ``Entropy discrepancy and total derivatives in trace anomaly,''
  arXiv:1412.0452 [hep-th].

\bibitem{Miao}
  R.~X.~Miao and W.~z.~Guo,
  ``Holographic Entanglement Entropy for the Most General Higher Derivative Gravity,''
  JHEP {\bf 1508}, 031 (2015)
  [arXiv:1411.5579 [hep-th]].

\bibitem{Maldacena:1997re} 
  J.~M.~Maldacena,
  ``The Large N limit of superconformal field theories and supergravity,''
  Int.\ J.\ Theor.\ Phys.\  {\bf 38}, 1113 (1999)
  [Adv.\ Theor.\ Math.\ Phys.\  {\bf 2}, 231 (1998)]
  [hep-th/9711200].

\bibitem{Gubser:1998bc} 
  S.~S.~Gubser, I.~R.~Klebanov and A.~M.~Polyakov,
  ``Gauge theory correlators from noncritical string theory,''
  Phys.\ Lett.\ B {\bf 428}, 105 (1998)
  [hep-th/9802109].

\bibitem{Witten:1998qj} 
  E.~Witten,
  ``Anti-de Sitter space and holography,''
  Adv.\ Theor.\ Math.\ Phys.\  {\bf 2}, 253 (1998)
  [hep-th/9802150].

\bibitem{Camps:2014voa} 
  J.~Camps and W.~R.~Kelly,
  ``Generalized gravitational entropy without replica symmetry,''
  JHEP {\bf 1503}, 061 (2015)
  doi:10.1007/JHEP03(2015)061
  [arXiv:1412.4093 [hep-th]].

\bibitem{Dyer:2008hb} 
  E.~Dyer and K.~Hinterbichler,
  ``Boundary Terms, Variational Principles and Higher Derivative Modified Gravity,''
  Phys.\ Rev.\ D {\bf 79}, 024028 (2009)
  [arXiv:0809.4033 [gr-qc]].

\bibitem{Lovelock:1971yv} 
  D.~Lovelock,
  ``The Einstein tensor and its generalizations,''
  J.\ Math.\ Phys.\  {\bf 12}, 498 (1971).

\bibitem{Lovelock:1970} 
  D.~Lovelock,
  ``Divergence-free tensorial concomitants,''
  aequationes mathematicae {\bf 4} (1970), no.~1-2 127--138.

\bibitem{Jacobson:1993xs} 
  T.~Jacobson and R.~C.~Myers,
  ``Black hole entropy and higher curvature interactions,''
  Phys.\ Rev.\ Lett.\  {\bf 70}, 3684 (1993)
  [hep-th/9305016].


  \bibitem{Miao2} 
  R.~X.~Miao,
  ``Universal Terms of Entanglement Entropy for 6d CFTs,''
  JHEP {\bf 1510}, 049 (2015)
  [arXiv:1503.05538 [hep-th]].

\bibitem{Miao1}
  Y.~Huang and R.~X.~Miao,
  ``A note on the resolution of the entropy discrepancy,''
  Phys.\ Lett.\ B {\bf 749}, 489 (2015)
  [arXiv:1504.02301 [hep-th]].

\bibitem{Theisen}
  A.~Schwimmer and S.~Theisen,
  ``Entanglement Entropy, Trace Anomalies and Holography,''
  Nucl.\ Phys.\ B {\bf 801}, 1 (2008)
  [arXiv:0802.1017 [hep-th]].

\bibitem{Henningson}
  M.~Henningson and K.~Skenderis,
  ``The Holographic Weyl anomaly,''
  JHEP {\bf 9807}, 023 (1998)
  [hep-th/9806087].

\bibitem{Miao3}
  R.~X.~Miao,
  ``A Note on Holographic Weyl Anomaly and Entanglement Entropy,''
  Class.\ Quant.\ Grav.\  {\bf 31}, 065009 (2014)
  [arXiv:1309.0211 [hep-th]].

\bibitem{Solodukhin:2008dh} 
  S.~N.~Solodukhin,
  ``Entanglement entropy, conformal invariance and extrinsic geometry,''
  Phys.\ Lett.\ B {\bf 665}, 305 (2008)
  [arXiv:0802.3117 [hep-th]].

\bibitem{Hung}
  L.~Y.~Hung, R.~C.~Myers and M.~Smolkin,
  ``On Holographic Entanglement Entropy and Higher Curvature Gravity,''
  JHEP {\bf 1104}, 025 (2011)
  [arXiv:1101.5813 [hep-th]].
 
\bibitem{Myers}
  R.~C.~Myers and A.~Sinha,
  ``Holographic c-theorems in arbitrary dimensions,''
  JHEP {\bf 1101}, 125 (2011)
  [arXiv:1011.5819 [hep-th]].
  


\bibitem{Bueno:2015rda} 
P.~Bueno, R.~C.~Myers and W.~Witczak-Krempa,
``Universality of corner entanglement in conformal field theories,''
Phys.\ Rev.\ Lett.\  {\bf 115}, no. 2, 021602 (2015)
[arXiv:1505.04804 [hep-th]].

\bibitem{Faulkner:2015csl} 
  T.~Faulkner, R.~G.~Leigh and O.~Parrikar,
  ``Shape Dependence of Entanglement Entropy in Conformal Field Theories,''
  arXiv:1511.05179 [hep-th].

\bibitem{Bueno:2015xda} 
P.~Bueno and R.~C.~Myers,
``Corner contributions to holographic entanglement entropy,''
JHEP {\bf 1508}, 068 (2015)
[arXiv:1505.07842 [hep-th]].



\bibitem{Miao:2015dua} 
R.~X.~Miao,
``A holographic proof of the universality of corner entanglement for CFTs,''
JHEP {\bf 1510}, 038 (2015)
[arXiv:1507.06283 [hep-th]].

\bibitem{Bueno:2015lza} 
P.~Bueno and R.~C.~Myers,
``Universal entanglement for higher dimensional cones,''
arXiv:1508.00587 [hep-th].


\bibitem{Wall1}
  S.~Sarkar and A.~C.~Wall,
  ``Generalized second law at linear order for actions that are functions of Lovelock densities,''
  Phys.\ Rev.\ D {\bf 88}, 044017 (2013)
  [arXiv:1306.1623 [gr-qc]].

\bibitem{Wall2}
  S.~Bhattacharjee, S.~Sarkar and A.~C.~Wall,
  ``Holographic entropy increases in quadratic curvature gravity,''
  Phys.\ Rev.\ D {\bf 92}, no. 6, 064006 (2015)
  [arXiv:1504.04706 [gr-qc]].

\bibitem{Wall3}
  A.~C.~Wall,
  ``A Second Law for Higher Curvature Gravity,''
  arXiv:1504.08040 [gr-qc].
  
\bibitem{Bhattacharjee:2015qaa} 
S.~Bhattacharjee, A.~Bhattacharyya, S.~Sarkar and A.~Sinha,
``Entropy functionals and c-theorems from the second law,''
arXiv:1508.01658 [hep-th].

\end{thebibliography}
\end{document}